\documentclass[11pt]{article}
\input epsf.sty
\usepackage{epsfig} 
\newcommand{\beq}{\begin{eqnarray}}
\newcommand{\eeq}{\end{eqnarray}}
\newcommand{\be}{\begin{equation}}
\newcommand{\ee}{\end{equation}}
\newcommand{\sectiono}[1]{\section{#1}\setcounter{equation}{0}}
\newcommand{\Tr}{{\rm{Tr}}}

\begin{document}

{}~ \hfill\vbox{\hbox{hep-th/0503215}

\hbox{HUTP-05/A0014}

 }\break

\vskip 1cm

\begin{center}
\Large{\bf Long strings condensation and FZZT branes}

\vspace{0.6cm}

\vspace{20mm}

\large{Davide Gaiotto$^{a}$}

\vspace{10mm}

\large{\em $^{a}$ Jefferson Physical Laboratory, Harvard
University,}

\vspace{0.2cm} \normalsize{\em Cambridge, MA 01238, USA}

\vspace{0.4cm}

\end{center}

\vspace{10mm}

\begin{abstract}

\bigskip
We propose a matrix model description of extended D-branes in 2D noncritical string
\medskip

\end{abstract}

\newpage

\tableofcontents

\section{Introduction }

The singlet sector of large $N$ gauged quantum mechanics have long
been given an interpretation as an exact description of $c=1$
non-critical string theory. The closed string field theory
describes the collective motion of the matrix
eigenvalues\cite{Klebanov:1991qa,Ginsparg:1993is,DiFrancesco:1993nw,David:1990sk,Moore:1991zv,DiFrancesco:1991ud}.
The non-singlet sectors of the same quantum mechanics have long
resisted such a detailed interpretation. The recent work
\cite{Maldacena:2005hi} interprets the most basic non-singlet
sector, the adjoint, as a sector of $c=1$ string
theory in which one macroscopic fundamental string (long string) is stretched in
space from the weak-coupling infinity and folded back.

The turning point of this long string (the tip) is pulled back by
the string tension and repelled by the tachyon wall, hence the
string tip follows a scattering trajectory coming in from weak
coupling infinity in the far past and disappearing back in the far
future. The adjoint sector of the quantum mechanics is shown to
describe a Fermi sea of free eigenvalues with an interacting
impurity, and the impurity dynamics is identified with the
dynamics of the tip of the folded string.

\cite{Maldacena:2005hi} makes a further identification
to obtain a worldsheet CFT description of the adjoint sector: the folded
string is seen as the appropriate limit of an open string mode
living on an FZZT brane with Neumann conditions in time. (see for
example
\cite{Fateev:2000ik,Teschner:2000md,Nakayama:2004vk,Kostov:2003uh}
for a definition and review of FZZT branes in noncritical string
theories)

While most D-branes in $c<1$ noncritical strings have been
understood nicely in terms of resolvents of the corresponding
matrix models
\cite{Seiberg:2004at,Klebanov:2003km,Seiberg:2003nm,Gaiotto:2003yb,Kutasov:2004fg,Hanada:2004im},
in $c=1$ these FZZT branes with Neumann boundary conditions in
time have until now resisted a matrix model interpretation. These
D-branes are extended in time and space, and support a massless
propagating degree of freedom (the open string tachyon) that can
considerably enrich the dynamics of the $c=1$ model. The open
string tachyon modes are kept away from the strong coupling region
both by the closed string tachyon wall $\mu e^{-\phi}$ and by an
open tachyon wall $\mu_B e^{-\phi_o}$

\begin{figure}
\centering \epsfig{file=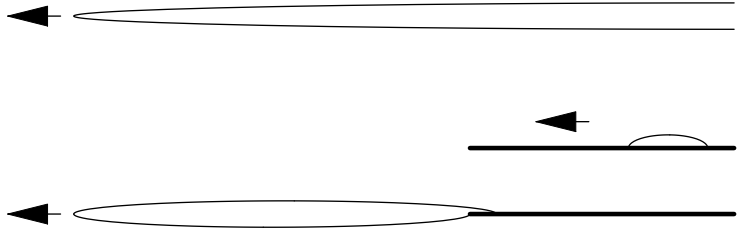, width=4in} \caption{A
stretched string is the limit of a very energetic open string mode
on a FZZT brane. \label{stretch}}
\end{figure}

In \cite{Maldacena:2005hi} the folded string is related to a very
energetic open string attached to an FZZT brane of very large
boundary cosmological constant $\mu_B$. The large open tachyon
wall $\mu_B e^{-\phi}$ pins the string endpoints far in the weak
coupling region at $\log \mu_B$, but the large energy allows the
open string bulk to to stretch macroscopically towards the strong
coupling region. The stretched string limit is reached by keeping
$E_{open}-\log \mu_B$ constant while sending $\mu_B$ to infinity,
so to keep finite the energy available to the string tip's motion.
(see Figure \ref{stretch})

If one keeps $\mu_B$ large but finite, an observer far enough in
the weak coupling region to see the FZZT brane will interprete the scattering process
 as a coherent pulse of open string radiation coming in and
condensing into a long stretched string. After a while the stretched
string will then decay back to open string radiation.

We observe here an obvious similarity between this scenario and
the description of unstable ZZ branes in the usual matrix model:
consider a very energetic close string mode travelling in a
background with very large bulk cosmological constant $\mu$. If
$E_{closed}-\log \mu$ is finite there will be a large amount of
time for which the closed string mode will admit a matrix model
description as one (or few) eigenvalues detached from the rest of
the Fermi sea, i.e. a decaying ZZ brane with a $\Lambda \cosh X^0$
tachyon deformation turned on the worldvolume.
\cite{Klebanov:2003km,McGreevy:2003kb,deBoer:2003hd,McGreevy:2003ep}

\begin{figure}
\centering \epsfig{file=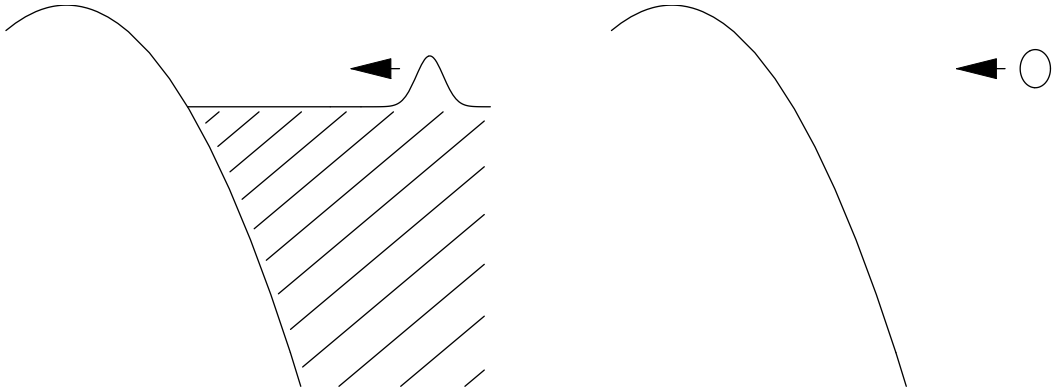, width=6in} \caption{A pulse of
closed string radiation and the result of $\mu \to \infty$
$E_{cl}-\log \mu = const.$. \label{closed}}
\end{figure}

The limit $\mu \to \infty$, $E_{closed}-\log \mu$ finite drains
the Fermi sea and leaves a finite number of eigenvalues with a
finite distance from the top of the potential.(see Figure
\ref{closed}) It is then possible to recover a finite $\mu$ by
condensing infinitely many of such eigenvalues and producing again
a Fermi sea. This is the statement that $c=1$ noncritical string
theory can be recovered from the collective coordinates of
infinitely many condensed ZZ branes, and that condensation of ZZ
branes is equivalent to a renormalization of $\mu$

We want to follow a similar logic to renormalize the very large
$\mu_B$ relevant for the long string setting back to a finite
value, recovering the full open string tachyon dynamics from the
collective coordinates of a large number of long strings. Indeed
there are sectors of the matrix model labelled by certain higher
representations of $U(N)$ that can be used to describe several
stretched strings, i.e. several impurities in the Fermi sea, in
such a way that these impurities have Fermi statistics themselves.

\begin{figure}
\centering \epsfig{file=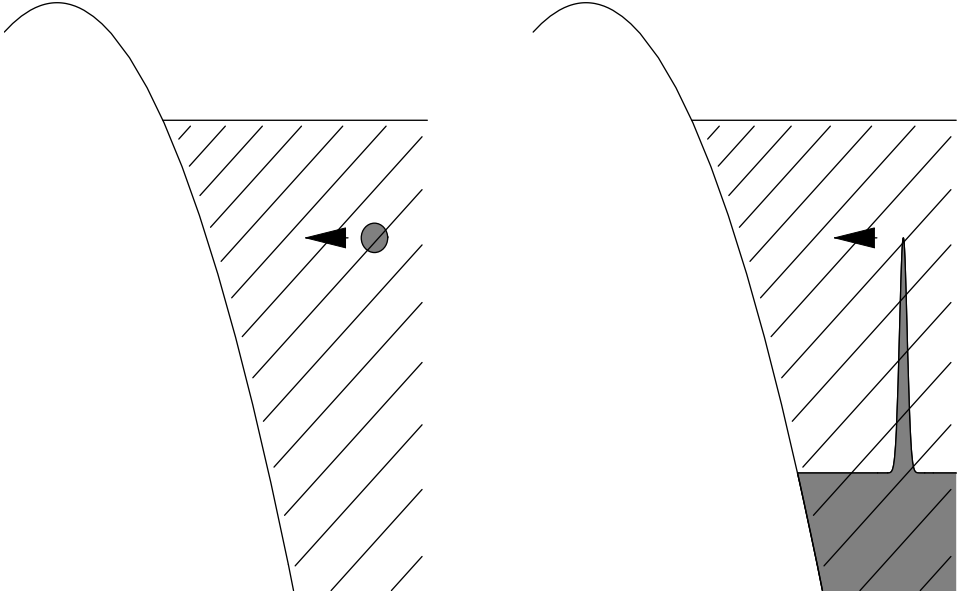, width=5in} \caption{A single
impurity in the Fermi sea and a pulse of open string radiation are
related by $\mu_B \to \infty$ $E_{op}-\log \mu_B = const.$.
\label{open}}
\end{figure}

By analogy with the closed string case, we propose the following
statement: the dynamics of an FZZT brane of finite $\mu_B$ in
$c=1$ noncritical strings is recovered by condensation of an
infinite number of long strings, and is encoded in sectors of the
matrix model corresponding to representations with a very large
number of boxes (impurities). We intend to show that in presence
of a chemical potential for the number of impurities a new Fermi
sea of impurities arises, whose collective fluctuations encode the
open string tachyon living on the FZZT brane. Stretched strings
attached to a finite $\mu_B$ FZZT brane will correspond to single
impurities well above the impurity Fermi sea. (see Figure
\ref{open})

The presence of a Fermi level for impurities implies that the energy of one impurity
above the vacuum state is finite (the extra impurity cannot have an energy lower that the Fermi sea)
This corresponds to the fact that the long string doesn't need to stretch to infinity,
but only to the tip of the FZZT brane.

The results in \cite{Maldacena:2005hi} cover the case of $0A$
and $0B$ $\hat c=1$ as well. A similar long string condensation process
as the one described here could be used to understand Neumann FZZT branes in those
non-critical superstring models.

The detailed organization of the paper can be gleaned from the
table of contents. In section two contains the main results of
this note: we will review some properties of non-singlet sectors,
give our precise matrix model prescription for the FZZT brane and
study the degrees of freedom of the system in the asymptotic
region. In section three we will compare the modified matrix model
action with the OSFT on a large number of decaying ZZ branes and
one FZZT brane. Section four contains some observations about the
consequences of our proposal in the theory with compact Euclidean
time. In section five we will give a proposal for the matrix model
description of an arbitrary number of FZZT branes in the $c=1$
model and attempt to give a rough comparison with the expected
target space dynamics.

\sectiono{Non-singlet sectors and impurities}

We want to propose an exact description of extended D-branes in
$c=1$ noncritical string theory. We will show that the sea of eigenvalues
in the sector of Matrix Quantum Mechanics
associated with a condensate of long strings
can supports two independent collective degrees of freedom,
to be identified with the closed string field and
the open string field living on a FZZT brane.

Matrix quantum mechanics is described by a very simple $SU(N)$
invariant action
\cite{Klebanov:1991qa,Ginsparg:1993is,Polchinski:1994mb}

\be S(M) = \int dt \, \Tr \left[ \dot M^2 - V(M) \right] \ee

The  restriction to the $SU(N)$ invariant sector of the quantum
mechanics is typically done by gauging it: $\dot M \to \dot M -
[A_0,M]$. This gauging has a straightforward interpretation when
the matrix model is taken to be the BSFT for a large number of
unstable ZZ branes. (see section 3 for more on this.)

The gauging removes the angular degrees of freedom, the
eigenvalues of M behave like free fermions in the potential
$V(\lambda)$ and form a Fermi sea in the potential well. An
appropriate double scaling limit, that makes $V(M)$ into an
upside-down harmonic oscillator $V(M) = -M^2$ while keeping the
Fermi level at finite distance from the top of the potential
yields the $c=1$ non-critical string theory. The only propagating
mode, the massless closed string tachyon, is mapped to
fluctuations of the fermi level by an appropriate bosonization
map.

Without gauging, the dynamics of the $N^2$ degrees of freedom in
$M$ still decomposes in infinitely many independent sectors, as
the large $SU(N)$ symmetry commutes with the Hamiltonian.
Different sectors are essentially labelled by the representations
of $SU(N)$ that appear in powers of the adjoint. More precisely
the change of variables

\be M = \Omega \; {\it diag} \{\lambda_1 ,\cdots, \lambda_N\} \;
\Omega^{-1} \ee

splits the phase space into eigenvalues and angular degrees of
freedom

\be \frac{R^N \otimes SU(N)/U(1)^{N-1}}{S_N} \ee

The quotient by $U(1)^{N-1}$ represents the freedom of acting from
the right on $\Omega$ with elements in the Cartan subalgebra,
$S_N$ the freedom to permute eigenvalues while acting at the same
time on $\Omega$ with the appropriate element of the Weyl group.

The Hilbert space $L^2\left(SU(N)/U(1)^{N-1}\right)$ admits a
natural complete orthogonal base given by Fourier theory on coset
spaces, on which the Hamiltonian block diagonal. We review in
appendix A the details of the decomposition, the well known result
is that

\be {\cal H} = \bigoplus_{\cal R} {\cal H}_{\cal R}^{d_{\cal R}}
\qquad {\cal H}_{\cal R} = \frac{L^2(R^N) \otimes \tilde{\cal
R}}{S_N} \ee

Here ${\cal R}$ are $SU(N)$ representations, $\tilde{\cal R}$ is
the subspace of zero weight vectors in ${\cal R}$, $d_{\cal R}$
the dimension of the representation and $S_N$ acts by simultaneous
permutation of the eigenvalues and of the corresponding Cartan
generators.

The Hamiltonian in any given sector is

\be \sum_i ( -\frac{\partial^2}{\partial \lambda_i^2} -
\lambda_i^2) + \sum_{i,j} \frac{1-\hat
P_{ij}}{(\lambda_i-\lambda_j)^2} \ee

$\hat P_{ij}$ acts on $\tilde {\cal R}$ as a basic transposition in the
Weyl group $S_N$, permuting the Cartan generators $i$ and $j$. \cite{Minahan:1993mv,Minahan:1992ht}

Reasoning along the lines of \cite{Maldacena:2005hi}, when ${\cal
R}$ is the adjoint representation (more properly adjoint plus
singlet, i.e. fundamental times anti-fundamental), $\tilde {\cal R}$ is the span of $N$ diagonal vectors
$|n\rangle \otimes |\bar n\rangle$, $n= 1 \cdots N$. The
Hamiltonian describes the coupling of the $N$ eigenvalues to a
discrete degree of freedom $n$, that acts as a label whose effect
is to distinguish one of the eigenvalues from the others. The
remaining $N-1$ eigenvalues do not interact with each other, but
only with the labelled $n-th$ eigenvalue. The interaction is the
sum of a repulsive term and an exchange term, that allows the
label ("impurity") to hop to a nearby eigenvalue.

In the double scaling limit one is left with the coupled system of
the Fermi sea plus impurity, and the impurity is identified with
the tip of the stretched string. \cite{Maldacena:2005hi} studies
the scattering phase shifts for the impurity, in the natural
approximation that ignores the back-reaction of the impurity on
the second-quantized Fermi sea.

We propose that the sectors relevant to stretched string
condensation are the ones labelled by the representation
${\cal R}_k = (N)^{\wedge k} \otimes (\bar N)^{\wedge k}$, where $(N)$ is the
fundamental, $(\bar N)$ the anti-fundamental of $SU(N)$. The zero
weight subspace $\tilde {\cal R}_k$ is span by ${N \choose k}$ vectors labelled by a
multi-index $I=\{i_1,\cdots,i_k\}$, $i_1<i_2<\cdots<i_k$.

These discrete degrees of freedom are equivalent to an assignment
of $k$ indistinguishable labels marking $k$ of the eigenvalues.
These $k$ eigenvalues interact as before with the remaining $N-k$
but do not interact between themselves. Following the analogy,
after double scaling limit it is natural to identify these $k$
impurities in the Fermi sea with $k$ stretched strings, that we
claim are attached to a single FZZT brane with very large $\mu_B$.

If there were no interaction terms in the Hamiltonian the system
would be equivalent to two sets of independent fermions in
original potential. It would be natural to send both $N-k$ and $k$
to infinity in order to describe a condensation of the long
strings. The chemical potential of the impurity Fermi sea would be
naturally identified with a finite $\mu_B$ for the FZZT brane.

While a single-particle Fermi sea description is available exactly
in the singlet sector and approximately in the small $k$ sectors,
this double scaling limit procedure in $k$ may seem foolhardy, as it
is far from obvious that for large $N-k$ and $k$ this highly
interacting (but integrable! \cite{Minahan:1993mv,Minahan:1992ht})
spin-Calogero model will admit any such description.

On the other hand there is a simple canonical transformation in
the original matrix model that makes evident the integrable
structure of the problem and shows that this picture of two Fermi
surfaces for two kind of (quasi-)particles is indeed appropriate.

Consider $P$, the matrix momentum conjugate to $M$, and make the
hermitian combinations $A^{\pm} = P \pm M$.

The Hamiltonian is naturally written as $\Tr \{A^{+},A^{-}\}$. It
is appropriate to make a change of base from wavefunctions of $M$
to wavefunctions of, say, $A^+$. The power of this change of
coordinates is the fact that diagonalization of $A^+ = \Omega^+
{\it diag}\{a^+_i\} \bar \Omega^+$ reorganizes each sector subtly,
so that the eigenvalues $a^+_i$ become free fermions, with
Hamiltonian

\be \sum_i \{a^+_i,-\imath \frac{\partial}{\partial a^+_i}\} \ee

and no two-particle interaction terms!

In the singlet sectors the changes of variable between $M$, $A^+$
and $A^-$ "commute" with the diagonalization, and translate
directly in the corresponding change of variables between
$\lambda_i$ and $a^\pm_i$. In a generic sector the matrix change
of variables maps each sector of the angular decomposition of the
Hilbert space for one of the variables into the corresponding
sector for the other variables. The change of base mixes in a
complicated way the eigenvalue degrees of freedom with the
discrete ones in $\tilde {\cal R}$, so that eigenvalues in one
description become quasi-particles in another. See appendix B for
more details on this.

The time evolution of $a^{\pm}$ is rescaling by $e^{\pm t}$, as
the Hamiltonian is the dilatation operator for $a^{\pm}$. In the
singlet sector the variables $\tau^+ =\log a^+$ and $\tau^- = \log
a^-$ are appropriate to describe the asymptotic outgoing and
ingoing regions, making the relativistic nature of the motion
obvious. Fourier transform of eigenfunctions from $a^-$ to $a^+$
is the traditional swift way to get the single fermion phase
shifts.

The energy eigenfunctions for the system in a generic sector can
be written down directly:

\be \psi^+_J(a^+_n, \epsilon_n) = \sum_{\sigma \in S_N} \epsilon(\sigma)
c_{\sigma(J)} \prod (a^+_{\sigma(n)})^{\imath \epsilon_n -
\frac{1}{2}} \ee

Scattering requires convolution with an integral kernel
$G^K_J(a^-_m,a^+_n)$ that can be computed by evaluating certain
Izkinson-Zuber-like $SU(N)$ integrals.

\be \psi^-_J(a^-_m) = \int \prod_k da^+_k \sum_{K \in \tilde {\cal
R}}G^K_J(a^-_m,a^+_n) \psi^+_K(a^+_n) \ee

It would be interesting to pursue this path to evaluate the
scattering phases of \cite{Maldacena:2005hi} on the whole energy
spectrum. Less ambitiously we just remark that in the sector
corresponding to the representations ${\cal R}_k$ the system is equivalent to two
independent sets of $N-k$ and $k$ free fermions. In fact the
internal degrees of freedom in this sector are essentially
equivalent to the choice of a subset of $k$ out of $N$
indistinguishable eigenvalues. One can trade this choice for a
reduction of the permutation symmetry from $S_N$ to $S_{N-k}
\otimes S_k$, and describe the system in terms of two different
species of fermionic eigenvalues.

To second-quantize the system, consider a direct sum of sectors
with different $k$

\be {\cal R}_{\it Fock} = \sum_{k = 0}^N (N)^{\wedge k} \otimes
(\bar N)^{\wedge k} \ee.

The sector ${\cal R}_{\it Fock}$ of matrix quantum mechanics
describes a system of two kind of fermions, or a system of $N$
fermions with an $SU(2)$ isospin
\cite{Minahan:1992ht,Minahan:1993mv,Maldacena:2005hi}. One
can then naturally define two interacting Fermi fields
$\Psi_i(\lambda)$ that creates/destroys eigenvalues of the two
kinds, changing the values of $N$ and $k$.

A non-trivial field redefinition exists to map these fields into corresponding
quasi-particle free Fermi fields $\Psi_i^\pm(a^\pm)$, built along the lines of the finite $N$ case.
Ingoing and outgoing vacua can be defined in a straightforward fashion, with
appropriate Fermi levels for the two species.

The story then follows closely the description in the singlet
sector, with twice as many fields. In terms of $\tau^\pm = \log
a^\pm$ these free Fermi fields represent free massless
relativistic fermions. The standard bosonization map can convert
them to two massless chiral bosonic fields
$\Phi_i^\pm(\tau^\pm,t)$.

We propose then to associate the closed and open string tachyons
with these two collective fields. We leave the precise
identification, that is the choice of appropriate two-by-two leg
pole matrices, to future work. These leg-pole coefficients should
be fixed by comparison of the simplest $1 \to 1$ scattering
amplitudes in the matrix model against the corresponding bulk two
point function on the sphere for the closed string tachyon, the
boundary two point function on the disk for the open string
tachyon and the mixed bulk to boundary two point function on the
disk.

After the leg poles have been fixed appropriately, higher
correlation functions may be compared.

A second kind of comparison would involve a careful study of the
boundary equivalent of the ground ring (see for example
\cite{Witten:1991zd,Kostov:2004cd}) and comparison with the symmetry algebra of
this $SU(2)$ spin-Calogero system, that extends the $A_\infty$ of the singlet sector.

The presence of an $SU(2)$ symmetry between the two species of
eigenvalues is an interesting and unexpected fact from the point
of view of the target space physics. As long as $\mu_B^2 \gg \mu$ it is natural
to identify excitations of the upper Fermi sea with the closed strings, that can reach
as deep as $- \log \mu$ in the strong coupling region, and excitations of the lower Fermi sea
with open strings, that reflect back on the open tachyon wall already at $-\log \mu_B$.
It is possible to lower the boundary cosmological constant until the two Fermi
levels become comparable, but then as they overcome each other
and become well separated again, the upper Fermi sea should again
be identified with closed strings. It appears that the moduli
space for $\frac{\mu_B^2}{\mu}$ should be an half line, starting from a
point of extended $SU(2)$ symmetry. It would be interesting to
verify if the special properties of this point are manifest in the
worldsheet BCFT description of the model, for large but comparable $\mu$ and
$\mu_B$.

A second important issue is to understand the relation between the incoming vacuum
defined by two Fermi seas of incoming quasi-particles and
the outgoing vacuum defined by the two Fermi seas of outgoing
quasi-particles. While it is possible that they are mapped onto each other
by the second-quantized version of the matrix Fourier transform, it would be a fairly non-trivial fact.
It is more probable that the incoming vacuum for a FZZT brane will
map into a non-trivial state with open and closed radiation in the outgoing region.
This should be checked against the possibility of pair creation in the background of an FZZT brane.

\sectiono{FZZT - ZZ open strings}

There is an alternative, suggestive way to describe a
second-quantized ensemble of representations $\bigoplus_k
(N^{\wedge k} \otimes \bar N^{\wedge k})$. Consider an extended
gauged quantum mechanics, with action

\be S(M,u,\bar u) = \int dt \, \Tr \left[ (D_0 M)^2 + M^2 \right]
+ u^\dagger D_0 u + \bar u^\dagger D_0 \bar u\ee

We added two fermionic oscillators, $u$ in the fundamental of
$U(N)$ and $\bar u$ in the anti-fundamental. The gauging reduces
the Hilbert space to the global singlet representation, that is

\be {\cal H}_s = \bigoplus_k {\cal H}_u^{\bar N^{\wedge k}}
\otimes {\cal H}_{\bar u}^{N^{\wedge k}} \otimes {\cal
H}_M^{N^{\wedge k} \otimes \bar N^{\wedge k}} \ee

In other words the relevant sectors of the Matrix quantum
mechanics appear in the global singlet paired up with the
appropriate sectors of the $u,\bar u$ system. At first sight there
is a certain over-counting, but after diagonalization of $M$ it is
easy to check that each desired sector $k$ appears exactly once:
the residual $U(1)^N$ invariance forces the $u$ Hilbert
space to be the span of $\prod (u^\dagger_i \bar u^\dagger_i)
|0\rangle$, where $|0\rangle$ is killed by $u_i, \bar u_i$. Each
base vector gives corresponds to the choice of a subset of
$\{1,\cdots,N\}$. The system is the same as the one we propose for
the long string condensation.

This offers an alternative language to define the collective fields for the two Fermi surfaces  a la
Das-Jevicki\cite{Das:1990ka} roughly as

\be \phi(x,t) = \Tr \delta(x-M(t)) \qquad \phi_o(x,t) = u^\dagger
\, \delta(x-M(t)) \, u \ee

We propose to identify this gauged quantum mechanics with the BSFT
for the degrees of freedom of $N$ unstable ZZ branes and one FZZT
brane. The matrix $M$ should represent the open string tachyon on
the decaying ZZ branes, and $u, \bar u$ the open strings stretched
between FZZT and ZZ branes. The quantum mechanics is gauged
because of the gauge field living on the unstable ZZ branes. In
$c<1$ noncritical strings a similar idea to obtain a description
of FZZT branes by integrating away open string modes between the
FZZT and the ZZ branes has been successfully implemented.
\cite{Maldacena:2004sn}

An important difference with the $c<1$ case is that here we do not
require any non-trivial interaction between $M$ and $u, \bar u$
besides the indirect one mediated by the gauge field. This fact
allows the gauge-fixed action to be quadratic and solvable. A
possible way to argue for the absence of terms like $u^\dagger
F(M) u$ from the action is that the double scaling limit procedure
will generically scale away the functional dependence $F(M)$ to
leave a constant mass term for $u$, to be reabsorbed in the
definition of the chemical potential $\mu_B$. See also
\cite{Takayanagi:2005tq} for a possible alternative to the double
scaling limit.

\sectiono{FZZT branes and T-duality}

The thermal partition function in a given sector of the matrix
quantum mechanics can be computed in a straightforward way and has
been given an interpretation in the context of of $c=1$ in
compactified Euclidean time. (see for example
\cite{Kazakov:2000pm,Kazakov:2001pj,Alexandrov:2003ut,Gross:1990md})

The projection to a given representation can be implemented
directly in the trace of the thermal green function as

\be {\cal Z_R} = \int dM \,d\Omega \, G^E_\beta(M,\Omega M
\Omega^{-1}) \,Tr_{\cal R}(\Omega) \ee

Correlators of closed string winding modes in the Euclidean time
direction are computed by insertion of appropriate traces
$\frac{1}{n}\Tr \Omega^n$ in the singlet partition function. Hence
the twisted thermal partition function in a given sector ${\cal
R}$ corresponds to the computation of a winding-modes correlator.
What is the correlator that correspond to our long strings
condensation process? We need to sum over all the sectors, weighed
by the chemical potential for the stretched strings: $e^{-2 \mu_B
k} {\cal Z}_{N^{\wedge k} \otimes \bar N^{\wedge k}}$.

The corresponding sum of traces of exterior powers of the
fundamental can be traded for the insertion of a determinant.

\be {\cal Z}_{FZZT}(\mu_B) = \int dM \, d\Omega \,
G^E_\beta(M,\Omega M \Omega^{-1}) \, \det(1+ e^{-\mu_B}\Omega)\,
\det(1+ e^{-\mu_B}\Omega^{-1})\ee

This makes the relation between Neumann FTTZ branes and a
condensate of long strings evident: the insertion of the
determinant can regarded as the insertion of an exponentiated
winding loop operator $\exp(\Tr \log (1+e^{-\mu_B}\Omega) + \Tr
\log (1+e^{-\mu_B}\Omega^{-1}))$ extended along the circle
direction, and interpreted much along the lines of what has been
done in $c<1$ in terms of the winding integrable hierarchy.

A different kind of FZZT brane object that has been given an
interpretation in the matrix model is the FZZT brane istanton with
Dirichlet boundary conditions in time. This has been identified
\cite{Martinec:2003ka,Takayanagi:2004jz} with a basic matrix
resolvent $\det(z-M(t_0))$ placed at a particular time. In
euclidean time the various images along the circle would add up to
an appropriate momentum loop operator.

\beq \prod_n \det(z-M(2 \pi R n)) &=& \exp \left[\sum_n \Tr \log
\left(z-M(2 \pi R n)\right)\right] = \\  \exp \left[\sum_n \int d
t_E e^{i n t_E}\Tr \log (z-M(t_E)) \right] & \simeq & \exp
\left(\sum_n c_n T_n + \bar c_n T_{-n} \right)\eeq

T-duality acts by exchanging loop operators made of winding and
the momentum modes, and appropriately exchanges the FZZT brane
with Dirichlet and Neumann conditions in time.

\sectiono{$U(n)$ extension}

The comparison with the FTTZ-ZZ BSFT offers an immediate
generalization of our proposal that may describe the degrees of
freedom living on $n$ FZZT branes.

\be S(M,U,\bar U) = \int dt \Tr \left[ (D_0 M)^2 + M^2 \right] +
\Tr \left[ U^\dagger D_0 U \right] + \Tr \left[ \bar U^\dagger D_0
\bar U \right]\ee

Here the $Us$ are appropriate $N$ by $n$ matrices of fermionic
oscillators, corresponding to the open degrees of freedom between
$n$ FZZT and $N$ unstable ZZ branes. Note that we do not gauge the
$U(n)$ symmetry of the model, as the presence of a gauge field on
FZZT branes side is questionable. One can construct a vertex
operator that roughly corresponds to it, but it behaves much like
the discrete states in the closed string sector. This matter is of
considerable interest and is related to the issue of generalizing
the $A_\infty$ symmetry algebra of the closed string theory to the
open-closed string theory.

After diagonalization of $M$ the fermionic variables $U$ produce
states of the form $\prod (U^\dagger_{i,m} \bar U^\dagger_{i,m`})
|0\rangle$. These may be thought as an assignment of $n^2$ kind of
labels to the eigenvalues, so one has $n^2$ types of impurities
that can move around the eigenvalue Fermi sea.

On the other hand we can add $n$ chemical potentials as masses for
the $U$ fermionic oscillators, and identify them with $n$ boundary
cosmological constants on the $n$ FZZT branes. It is natural to
associate the dynamics of the $n^2$ types of impurities with the
$n^2$ types of long strings attached to $n$ FZZT branes. The
situation is considerably more complicated than in the case of one
FZZT brane, and there are considerable subtleties to be understood
in the scaling limit. We leave that to future work.

\section*{Acknowledgments}

We are grateful to N. Maldacena for illuminating discussions and
suggestions. We are glad to acknowledge useful conversations with
I. Kostov, D. Shih, A.Strominger and T. Takayanagi. We want to thank the Center of Mathematical Sciences at
Zhejiang University, Hangzhou, for the gracious hospitality.

This work was supported by DOE grant DE-FG02-91ER40654.

\appendix

\sectiono{Decomposition in non-singlet sectors}

The Hilbert space of square integrable functions on a compact
group $L^2(G)$ has a natural complete orthogonal base given by the
matrix functions that represent $G$ on its various
representations.

\be f(g) = \sum_{{\cal R},{I,J}\in {\cal R}} f^{{\cal R} I}_J
D_{{\cal R} I}^J(g) \ee

It is an important theorem of group theory that these matrix
element functions are orthogonal for the Haar invariant measure on
$G$. A traditional example would be the base $D^j_{m
m'}(\theta,\phi,\psi)$ on $SU(2) = S^3$.

The natural action of $G_L * G_R$ on $L^2(G)$ maps to the action
on the indices $I,J$

\be f(g_L g g_R^{-1}) \to D_{{\cal R} K}^I(g_L) f^{{\cal R} K}_T
D_{{\cal R} J}^T(g_R^{-1}) \ee

To get a base for functions on a coset $G/H$ one just needs to
restrict the index $J$ to the $H$ invariant subspace. Again the
basic example is $SU(2)/U(1) = S^2$, with the spherical harmonics
$Y_{lm}(\theta,\phi) = D^l_{m,m'=0}(\theta,\phi,\psi)$.

$SU(N)$ invariance of the action implies that the Hamiltonian
commutes with the left $SU(N)$ action on $\Omega$, hence indices
${\cal R}$ and $I$ label independent dynamical sectors. The
Hilbert space is then decomposed to

\be {\cal H} = \bigoplus_{\cal R} {\cal H}_{\cal R}^{d(\cal R)}
\qquad {\cal H}_{\cal R} = \frac{L^2(R^N) \otimes \tilde{\cal
R}}{S_N} \ee

where $\tilde{\cal R}$ is the subspace of zero weight vectors in
${\cal R}$, and $S_N$ acts by permutation of the eigenvalues
combined with an action byu the corresponding Weyl group element.

The wavefunction in a given sector is simply

\be \Psi(M) = \sum_{J \in \tilde {\cal R}}
\frac{\psi_J(\lambda_n)}{\prod_{i>j}(\lambda_i - \lambda_j)}
D_{{\cal R} J}^I(\Omega) \ee

The Vandermonde factor comes as usual from the Jacobian of the
diagonalization.

\sectiono{Matrix fourier transform in non-singlet sectors}

In the paper we will need to understand the effect of changes of
variables from $M$ to $A^\pm = P \pm M$, where $P$ is the momentum
conjugate to M. These are all roughly of the form of a Matrix
Fourier transform.

\beq \Psi(M) &=& \int dA^\pm e^{\Tr \left[\frac{\imath}{2}(A^\pm)^2+ \imath M A^\pm \pm
\frac{\imath}{2}M^2 \right]} \Psi_\pm(A^\pm) \\
\Psi_-(A^-) &=& \int dA^+ e^{\Tr \left[ \imath A^+ A^- \right]}
\Psi_+(A^+) \eeq

We want to show that these transformations map a specific angular
sector of the Hilbert space for one set of wavefunctions, say
$\Psi_+(A^+)$, to the corresponding sector for the other set, say
$\Psi_-(A^-)$.

Starting from the result of the previous appendix

\be \Psi+(A^+) = \sum_{J \in \tilde {\cal R}}
\frac{\psi_J(a^+_n)}{\prod_{i>j}(a^+_i - a^+_j)} D_{{\cal R}
J}^I(\Omega^+) \ee

Inserting this into the integral, and diagonalize the integration
variables to get

\be \Psi_-(A^-) = \int \prod_k da^+_k d\Omega^+ e^{\Tr \left[
\imath \Omega^+ diag\{a^+_n\} \bar \Omega^+ A^- \right]} \sum_{J
\in \tilde {\cal R}} \psi_J(a^+_n)\prod_{i>j}(a^+_i - a^+_j)
D_{{\cal R} J}^I(\Omega^+) \ee

Diagonalizing $A^-$ and changing variables in the integral
$\Omega^+ = \Omega^- \Omega$ gives

\be \Psi_-(\Omega^- diag\{a^-_m\} \bar \Omega^-) = \int \prod_k
da^+_k d\Omega e^{\Tr \left[ \imath \Omega diag\{a^+_n\} \bar
\Omega diag\{a^-_m\} \right]} \sum_{J,K \in \tilde {\cal R}}
\psi_J(a^+_n)\prod_{i>j}(a^+_i - a^+_j) D_{{\cal R}
K}^I(\Omega^-)D_{{\cal R} J}^K(\Omega) \ee

Hence ${\cal H}^+_{{\cal R},I}$ is mapped to ${\cal H}^-_{{\cal
R},I}$ by the transformation

\be \psi^-_J(a^-_m) = \int \prod_k da^+_k \sum_{K \in \tilde {\cal
R}}G^K_J(a^-_m,a^+_n) \psi^+_K(a^+_n) \ee

where

\be G^K_J(a^-_m,a^+_n) = \int d\Omega e^{\Tr \left[ \imath \Omega
diag\{a^+_n\} \bar \Omega diag\{a^-_m\} \right]} \prod_{i>j}(a^+_i
- a^+_j)(a^-_i - a^-_j) D_{{\cal R} J}^K(\Omega) \ee

This integral Kernel can be computed with character expansion in a
manner similar to the Izkinson-Zuber integral, but that goes
beyond the scope of this note.

\begingroup\raggedright

\endgroup


\begin{thebibliography}{10}

\bibitem{Maldacena:2005hi}
  J.~Maldacena,
  ``Long strings in two dimensional string theory and non-singlets in the
  matrix model,''
  arXiv:hep-th/0503112.

\bibitem{Klebanov:2003km}
  I.~R.~Klebanov, J.~Maldacena and N.~Seiberg,
  ``D-brane decay in two-dimensional string theory,''
  JHEP {\bf 0307}, 045 (2003)
  [arXiv:hep-th/0305159].

\bibitem{Klebanov:1991qa}
  I.~R.~Klebanov,
  ``String theory in two-dimensions,''
  arXiv:hep-th/9108019.

\bibitem{McGreevy:2003kb}
  J.~McGreevy and H.~Verlinde,
  ``Strings from tachyons: The c = 1 matrix reloaded,''
  JHEP {\bf 0312}, 054 (2003)
  [arXiv:hep-th/0304224].

\bibitem{Takayanagi:2005tq}
  T.~Takayanagi and S.~Terashima,
  ``c=1 Matrix Model from String Field Theory,''
  arXiv:hep-th/0503184.

\bibitem{Ginsparg:1993is}
  P.~H.~Ginsparg and G.~W.~Moore,
  ``Lectures on 2-D gravity and 2-D string theory,''
  arXiv:hep-th/9304011.

\bibitem{DiFrancesco:1993nw}
  P.~Di Francesco, P.~H.~Ginsparg and J.~Zinn-Justin,
  ``2-D Gravity and random matrices,''
  Phys.\ Rept.\  {\bf 254}, 1 (1995)
  [arXiv:hep-th/9306153].

\bibitem{David:1990sk}
  F.~David,
  ``Phases Of The Large N Matrix Model And Nonperturbative Effects In 2-D
  Gravity,''
  Nucl.\ Phys.\ B {\bf 348}, 507 (1991).

\bibitem{Moore:1991zv}
  G.~W.~Moore, M.~R.~Plesser and S.~Ramgoolam,
  ``Exact S matrix for 2-D string theory,''
  Nucl.\ Phys.\ B {\bf 377}, 143 (1992)
  [arXiv:hep-th/9111035].

\bibitem{Polchinski:1994mb}
  J.~Polchinski,
  ``What is string theory?,''
  arXiv:hep-th/9411028.

\bibitem{DiFrancesco:1991ud}
  P.~Di Francesco and D.~Kutasov,
  ``World sheet and space-time physics in two-dimensional (Super)string
  theory,''
  Nucl.\ Phys.\ B {\bf 375}, 119 (1992)
  [arXiv:hep-th/9109005].

\bibitem{Fateev:2000ik}
  V.~Fateev, A.~B.~Zamolodchikov and A.~B.~Zamolodchikov,
  ``Boundary Liouville field theory. I: Boundary state and boundary  two-point
  function,''
  arXiv:hep-th/0001012.

\bibitem{Teschner:2000md}
  J.~Teschner,
  ``Remarks on Liouville theory with boundary,''
  arXiv:hep-th/0009138.

\bibitem{Minahan:1993mv}
  J.~A.~Minahan and A.~P.~Polychronakos,
  ``Interacting fermion systems from two-dimensional QCD,''
  Phys.\ Lett.\ B {\bf 326}, 288 (1994)
  [arXiv:hep-th/9309044].

\bibitem{Minahan:1992ht}
  J.~A.~Minahan and A.~P.~Polychronakos,
  ``Integrable systems for particles with internal degrees of freedom,''
  Phys.\ Lett.\ B {\bf 302}, 265 (1993)
  [arXiv:hep-th/9206046].

\bibitem{Zamolodchikov:1995aa}
  A.~B.~Zamolodchikov and A.~B.~Zamolodchikov,
  ``Structure constants and conformal bootstrap in Liouville field theory,''
  Nucl.\ Phys.\ B {\bf 477}, 577 (1996)
  [arXiv:hep-th/9506136].

\bibitem{Das:1990ka}
  S.~R.~Das and A.~Jevicki,
  ``String Field Theory And Physical Interpretation Of D = 1 Strings,''
  Mod.\ Phys.\ Lett.\ A {\bf 5}, 1639 (1990).

\bibitem{Maldacena:2004sn}
  J.~Maldacena, G.~W.~Moore, N.~Seiberg and D.~Shih,
  ``Exact vs. semiclassical target space of the minimal string,''
  JHEP {\bf 0410}, 020 (2004)
  [arXiv:hep-th/0408039].

\bibitem{Kazakov:2000pm}
  V.~Kazakov, I.~K.~Kostov and D.~Kutasov,
  ``A matrix model for the two-dimensional black hole,''
  Nucl.\ Phys.\ B {\bf 622}, 141 (2002)
  [arXiv:hep-th/0101011].

\bibitem{Kazakov:2001pj}
  V.~A.~Kazakov and A.~A.~Tseytlin,
  ``On free energy of 2-d black hole in bosonic string theory,''
  JHEP {\bf 0106}, 021 (2001)
  [arXiv:hep-th/0104138].

\bibitem{Alexandrov:2003ut}
  S.~Alexandrov,
  ``Matrix quantum mechanics and two-dimensional string theory in  non-trivial
  backgrounds,''
  arXiv:hep-th/0311273.

\bibitem{Gross:1990md}
  D.~J.~Gross and I.~R.~Klebanov,
  ``Vortices And The Nonsinglet Sector Of The C = 1 Matrix Model,''
  Nucl.\ Phys.\ B {\bf 354}, 459 (1991).

\bibitem{Takayanagi:2004jz}
  T.~Takayanagi,
  ``Notes on D-branes in 2D type 0 string theory,''
  JHEP {\bf 0405}, 063 (2004)
  [arXiv:hep-th/0402196].

\bibitem{Martinec:2003ka}
  E.~J.~Martinec,
  ``The annular report on non-critical string theory,''
  arXiv:hep-th/0305148.

\bibitem{Kostov:2004cd}
  I.~K.~Kostov,
  ``Boundary ground ring and disc correlation functions in Liouville quantum
  gravity,''
  arXiv:hep-th/0402098.

\bibitem{Nakayama:2004vk}
  Y.~Nakayama,
  ``Liouville field theory: A decade after the revolution,''
  Int.\ J.\ Mod.\ Phys.\ A {\bf 19}, 2771 (2004)
  [arXiv:hep-th/0402009].

\bibitem{Kostov:2003uh}
  I.~K.~Kostov, B.~Ponsot and D.~Serban,
  ``Boundary Liouville theory and 2D quantum gravity,''
  Nucl.\ Phys.\ B {\bf 683}, 309 (2004)
  [arXiv:hep-th/0307189].

\bibitem{Seiberg:2004at}
  N.~Seiberg and D.~Shih,
  ``Minimal string theory,''
  arXiv:hep-th/0409306.

\bibitem{Seiberg:2003nm}
  N.~Seiberg and D.~Shih,
  ``Branes, rings and matrix models in minimal (super)string theory,''
  JHEP {\bf 0402}, 021 (2004)
  [arXiv:hep-th/0312170].

\bibitem{Gaiotto:2003yb}
  D.~Gaiotto and L.~Rastelli,
  ``A paradigm of open/closed duality: Liouville D-branes and the  Kontsevich
  model,''
  arXiv:hep-th/0312196.

\bibitem{Kutasov:2004fg}
  D.~Kutasov, K.~Okuyama, J.~w.~Park, N.~Seiberg and D.~Shih,
  ``Annulus amplitudes and ZZ branes in minimal string theory,''
  JHEP {\bf 0408}, 026 (2004)
  [arXiv:hep-th/0406030].

\bibitem{Hanada:2004im}
  M.~Hanada, M.~Hayakawa, N.~Ishibashi, H.~Kawai, T.~Kuroki, Y.~Matsuo and T.~Tada,
  ``Loops versus matrices: The nonperturbative aspects of noncritical string,''
  Prog.\ Theor.\ Phys.\  {\bf 112}, 131 (2004)
  [arXiv:hep-th/0405076].

\bibitem{deBoer:2003hd}
  J.~de Boer, A.~Sinkovics, E.~Verlinde and J.~T.~Yee,
  ``String interactions in c = 1 matrix model,''
  JHEP {\bf 0403}, 023 (2004)
  [arXiv:hep-th/0312135].

\bibitem{McGreevy:2003ep}
  J.~McGreevy, J.~Teschner and H.~Verlinde,
  ``Classical and quantum D-branes in 2D string theory,''
  JHEP {\bf 0401}, 039 (2004)
  [arXiv:hep-th/0305194].

\bibitem{Witten:1991zd}
  E.~Witten,
  ``Ground ring of two-dimensional string theory,''
  Nucl.\ Phys.\ B {\bf 373}, 187 (1992)
  [arXiv:hep-th/9108004].




\end{thebibliography}
\end{document}